\documentclass[prb,aps,onecolumn,amsmath,amssymb,floatfix, nofootinbib]{revtex4}

\usepackage{graphicx}
\usepackage{dcolumn}
\usepackage{bm}
\usepackage{xcolor}
\usepackage{amsmath}
\usepackage{comment}
\usepackage{enumitem}
\usepackage{float}
\usepackage{caption}
\usepackage{subcaption}
\usepackage[utf8]{inputenc}
\usepackage{mathrsfs}

\newcommand{\ed}{\end{document}}
\newcommand{\beq}{\begin{equation}}
\newcommand{\eeq}{\end{equation}}
\newcommand{\beqa}{\begin{eqnarray}}
\newcommand{\eeqa}{\end{eqnarray}}
\newcommand{\bc}{\begin{center}}
\newcommand{\ec}{\end{center}}

\newcommand{\ba}{\begin{array}}
\newcommand{\ea}{\end{array}}

\begin{document}
\title{ Solving Superconducting Quantum Circuits in Dirac's Constraint Analysis Framework}

\author{Akshat Pandey\footnote{email: apandey.physics@gmail.com (corresponding author)}}
\email{apandey.physics@gmail.com }

\author{Subir Ghosh}%
 \email{subirghosh20@gmail.com}
\affiliation{Physics and Applied Mathematics Unit\\Indian Statistical Institute, \\ 203, Barrackpore Trunk Road, Kolkata 700108, India}

 \maketitle
{\it{The paper is dedicated to the memory of Professor Roman Jackiw, who apart from working in diverse branches of theoretical physics, has contributed significantly in quantization of constraint systems}}.
\vskip .5cm
{\bf{Abstract:}}
In this work  we exploit Dirac's Constraint Analysis (DCA)   in Hamiltonian formalism to study different types of Superconducting Quantum Circuits (SQC) in a {\it{unified}} way. The Lagrangian of a SQC reveals the constraints, that are classified in a Hamiltonian framework, such that redundant variables can be removed to  isolate the canonical degrees of freedom for subsequent quantization of the Dirac Brackets via a generalized Correspondence Principle. This purely algebraic approach makes the application of concepts such as graph theory, null vector, loop charge,\ etc that are in vogue, (each for a specific type of circuit), completely redundant.  The universal validity of DCA scheme in SQC, proposed by us, is demonstrated by correctly re-deriving existing results for different SQCs, obtained previously exploiting different formalisms each applicable for a specific SQC. Furthermore, we have also analysed and predicted   new results for a   generic  form of SQC - it will be  interesting to see its validation in an  explicit circuit implementation. 
\vskip .5cm

{\bf{Introduction:}} Quantum electromagnetic Circuit \cite{nn} is an ubiquitous element   of the major operational models of Quantum computing. It has 
 straightforward analogy with classical computers: wires connecting gates that in turn  manipulate qubits. The transformation performed by the gates are always  reversible, contrary to   classical computers. SQCs are essential elements for  Quantum Computation  \cite{qcomp} and designing of qubits have been   diversified in to other design variations, such as quantronium
\cite{quant}, transmon \cite{trans}, fluxonium \cite{flux}, and
“hybrid” \cite{hybrid} qubits, all of which are less sensitive to decoherence mechanisms with
 improved   performance. Quantum effects at the macroscopic level are manifested in the field of circuit quantum electrodynamics \cite{sqed} where SQC as qubits  interact strongly in a controlled way  with microwave photons.

 In this Letter, we exploit Dirac's Constraint Analysis (DCA) in  Hamiltonian framework \cite{dir} to solve specific models of Superconducting Quantum Circuits (SQC). we stress that this method is systematic, {\it{universally}} applicable and at the same time operationally simple. This formalism is widely applied in Quantum Mechanics and Quantum Field Theory and important works in the context of SQC are \cite{para,mart}.{\footnote{In \cite{mart} it is observed that Dirac's method can fail in certain models; however our view is that the reason for failure should be attributed to the models in question (that are pathological in nature and physically unacceptable, as admitted by the authors themselves) and not on Dirac's formalism.}} In the context of SQC we quote from  \cite{sym} "{\it{... number of degrees of freedom is not equal to the number of Josephson Junctions or Quantum Phase Slips; worse, the fluxes and charges across the different elements may not be conjugate pairs. Understanding how to even identify the dynamical degrees of freedom, let alone quantize the circuit, is an open problem"...}};  clearly  the most important question is identification of  Canonical Degrees Of Freedom (CDOF), (to implement quantization of the circuit). Different types of SQC seem to require different methods - {\it{eg}} graph theory  \cite{sym} or loop charges  \cite{dual}. Thus even though the above works are successful in quantizing the circuits the situation is not satisfactory due to the lack of a unified picture; one has to look for an appropriate scheme for a specific circuit.  In this perspective DCA is a unified approach that makes all these diverse sophisticated mathematical tools redundant.

SQCs, instead of microscopic elements ( qubits),   are constructed out of macroscopic entities such as  electrical (LC) oscillator;  the collective motion of electron passing  without friction  at low temperature  and  Josephson effect,  introducing   nonlinearity without  dissipation or dephasing,
 is described by  flux threading the inductor,  acting as  center-of-mass position in a mass-spring mechanical oscillator. Conventional inductor-capacitor-resistance (LCR) circuit with batteries has a Lagrangian description with DOF being magnetic fluxes or electric
charges. {\it{However, in presence of constraints, (i.e. relations between circuit variables),   DCA is essential for a correct Hamiltonian analysis.}} In SQC theory a Hamiltonian framework is favoured, (in which DCA  also applies), since it  yields  quantum commutators from classical Poisson brackets via Correspondence Principle. CDOFs are essential for the latter. Apart from  an unambiguous identification of CDOF, there are other advantages in DCA: (i) it can reveal additional symmetries (if present) in the form of  gauge invariance, giving a new flexibility in building physically equivalent circuits, (ii) it can be used to construct a possible Lagrangian, yielding an SQC for  preferred sets of CDOF.  In passing we note that Symplectic Geometry formulation has also been used \cite{sym1,sym}, but possibly it is more convenient and economical to follow the Faddeev-Jackiw symplectic framework \cite{fad}.

Let us finally emphasize the novelty of our scheme of analysing SQC. As is clearly evident from the above, so far in existing literature, there is no unified framework for treating widely varying forms of SQCs and the methods of analysis are specific to each type of SQC. It appears that even the identification of canonical DOFs, to be subsequently quantised, is a non-trivial issue. In this perspective our work assumes proper significance: (i) We have established the universal applicability of DCA for different types of SQC by   correctly re-deriving existing results {\it{in a unified way}} for different SQCs, obtained previously exploiting different formalisms each applicable for a specific SQC. (ii)  Furthermore, we have also analysed and predicted   new results for a   generic  form of SQC - it will be  interesting to see its validation in an  explicit circuit implementation. 
The following subsections consist of; a brief description of DCA as is relevant here, a pedagogical example,  analysis of Inductively Shunted SQC in DCA approach, analysis of Capacitively  Shunted SQC in DCA approach and  analysis and prediction of new results for  a generic SQC in DCA approach.
\\
 {\bf{DCA relevant to us:}}  A more detailed discussion on DCA is provided in the Appendix. For a generic Lagrangian $L(q_i,\dot{q}_i)$, canonical momenta $p_i=(\partial L)/(\partial \dot{q}_i)$
 satisfy  Poisson Brackets (PB) $\{q_i,p_j\}=\delta_{ij},~\{q_i,q_j\}=\{p_i,p_j\}=0 $. Relations 
\begin{equation}
\chi_k=p_k-f_k(q_j)\approx 0
 \label{d2}   
\end{equation}
without $\dot{q}_i$ are treated as constraints. Next the extended form Hamiltonian, taking into account the constraints, is given by $H_E=p_i\dot{q}_i-L+\alpha_k\chi_k $
with  arbitrary multipliers $\alpha _k$. Time persistence  $\dot \chi_k=\{\chi_k,H_E\}\approx 0$
can give rise to  new constraints or in some cases some of the $\alpha_k$ can be determined (thus not generating new constraints). Note that "$\approx$" is a weak equality, modulo constraints, and  can not be used directly, in contrast to "$=$" strong equality  that can be used directly on dynamical variables. This process will continue until no further constraints are generated. Once the full set of linearly independent constraints are determined, the First Class Constraints (FCC), (associated with gauge invariance),  $\psi_r$ obeying $\{\psi_r,\chi_k\}\approx 0$ for all $k$ are isolated. Remaining  Second Class Constraints (SCC) are used to compute the non-singular constraint matrix $\chi_{kl}=\{\chi_k,\chi_l\}$. $\text{SCC}=0$ can be used  strongly  to eliminate some  DOF provided  Dirac brackets (DB) are used,  for  generic  $A,B$, 
\begin{equation}
\{A,B\}_{DB}=\{A,B\}-\{A,\chi_r\}{\chi}_{rs}^{-1}\{\chi_s,B\}.
 \label{d5}   
\end{equation}
 Notice that, for any arbitrary variable $A$, $\{A,\chi_r\}_{DB}=0$  is consistent with $\chi_r=0$ so that each SCC $\chi_r$ can remove one  DOF. Thus finally one has the Hamiltonian comprising of a reduced number of variables and time evolution of $A$ is computed using DBs
\begin{equation}
\dot A=\{A,H\}_{DB}.
 \label{df4}   
\end{equation}
Since $\chi_{kl}$ is anti-symmetric and non-singular there will be an even number of SCCs.\\
Remaining FCCs  reveal gauge invariance in the model. The FCCs,  together with respective gauge fixing conditions, give rise to a further set of SCCs, further  DBs, that will further  reduce the number of DOF. \\
{\it{DOF count}}: Each SCC can remove one DOF in phase space and each FCC can remove two DOF in phase space (FCC and its  gauge fixing condition) leading to the formula, in phase space,
\begin{equation}
({\text{no. of physical DOF}})=({\text{total no.  of DOF}}) 
-(2\times ({\text{no. of FCC}})+1\times ({\text{no. of SCC}})).
 \label{dff4}   
\end{equation}
Freedom of the choice of gauge fixing condition allows  manifestly different but physically equivalent models. \\

{\bf{A toy model:}} As an instructive pedagogical example we solve a simple toy model as shown in figure 1 to demonstrate the (in)action of  passive nodes. \footnote{We thank the Referee for suggesting this example.} Following \cite{mart}, the corresponding Lagrangian is

\begin{equation}
    L = \frac{C_2}{2} ( \dot{\phi}_3)^2 + \frac{C_1}{2} (\dot{\phi}_2 - \dot{\phi}_3)^2 - \frac{1}{2 L_1} (\phi_1 )^2 - \frac{1}{2 L_2} (\phi_2 - \phi_1)^2. 
\end{equation}

In the terminology of \cite{sym} $\phi_i$ are three flux node variables (with $\phi_0=0$ being earthed). The corresponding momenta $P_i=\frac{\partial L}{\partial \dot{\phi}_i}~,~i=0,1,2,3;~\{\phi_i,P_{j}\}=\delta_{ij}$ are given by

\begin{equation}
P_1=0,~P_2 = C_1(\dot \phi_2-\dot\phi_3),~P_3=C_2\dot\phi_3-C_1(\dot \phi_2-\dot\phi_3).
    \label{2}
\end{equation}

Note that there is a single constraint $\chi_1\equiv P_1\approx 0$. The canonical Hamiltonian $H=P_i\dot\phi_i-L$ follows:
\begin{equation}
    H = \frac{C_2}{2} ( \dot{\phi}_3)^2 + \frac{C_1}{2} (\dot{\phi}_2 - \dot{\phi}_3)^2 + \frac{1}{2 L_1} (\phi_1 )^2 + \frac{1}{2 L_2} (\phi_2 - \phi_1)^2
       \end{equation}
which, upon using (\ref{2}) reduces to
\begin{equation}
    H_E=\frac{(P_2+P_3)^2}{2C_2}+\frac{(P_2)^2}{2C_1}+ \frac{1}{2 L_1} (\phi_1 )^2 + \frac{1}{2 L_2} (\phi_2 - \phi_1)^2 +\lambda\chi_1
    \label{n1}
\end{equation}
where the extended Hamiltonian reads $H_E=H+\lambda\chi_1$.

Following DCA, time persistence of the constraint $\chi_1$ generates a further constraint $\psi$, given by
\begin{equation}
\psi \equiv \dot\chi=\{\chi,H_E\}=\frac{\phi_1}{L_1}-\frac{(\phi_2-\phi_1)}{L_1}\approx 0.
    \label{n2}
\end{equation}
Since $\{\chi,\psi\}=1/(L_1+L_2)$ is non-zero, $\chi,\psi$ constitute an SCC pair in DCA terminology and we can terminate the chain of constraints as we will be using the constraints as strong equations to reduce the phase space and Dirac Brackets will generate the dynamics. Using $\psi=0$ we remove $\phi_1=(L_1\phi_2)/(L_1+L_2)$  and compute the reduced $H$,
\begin{equation}
H=\frac{(P_2+P_3)^2}{2C_2}+\frac{(P_2)^2}{2C_1}+ \frac{1}{2 (L_1+L_2)} (\phi_2 )^2 . 
    \label{n3}
\end{equation}
It is important to note that Dirac Brackets, induced by the SCC pair $\chi,\psi$, is identical to the earlier canonical Poisson Brackets. The Hamiltonian equations of motion are easily recovered,
\begin{equation}
\dot\phi_2=\frac{(P_2+P_3)}{C_2}+\frac{P_2}{C_1},~~\dot P_2=-\frac{\phi_2}{L_1+L_2},~~\dot P_3=0.
    \label{n4}
\end{equation}
Thus we recover the mode equation
\begin{equation}
\ddot\phi_2=-\frac{1}{(L_1+L_2)}{(\frac{1}{C_1}+\frac{1}{C_2}})\phi_2
    \label{n5}
\end{equation}
where the frequency is simply $\Omega=1/{\sqrt{L_{eq}C_{eq}}}$ with $L_{eq}=L_1+L_2$ and $1/C_{eq}=1/C_1 +1/C_2$. These simply are the equivalent inductance and capacitance for a series combination. In fact from the circuit in Figure 1, the above result seems quite obvious and applying DCA in this case might look like using a sledgehammer to crack a nut! However, the general DCA established that the passive nodes are redundant in the circuit dynamics and subsequent more complicated circuit examples will clearly show the utility of DCA in SQC. 

\begin{figure}[!h]
\centering
\includegraphics[scale=0.3]{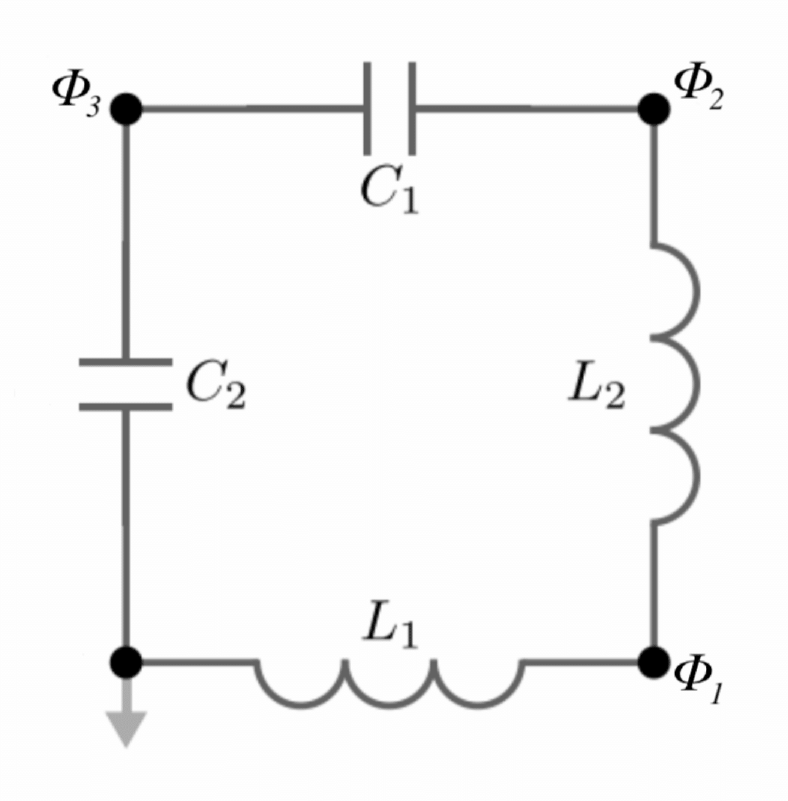}
\caption{A toy model circuit consisting of capacitors and conductors}
\label{}
\end{figure}

\begin{figure}
     \centering
     \begin{subfigure}[b]{0.25\textwidth}
         \centering
         \includegraphics[width=\textwidth]{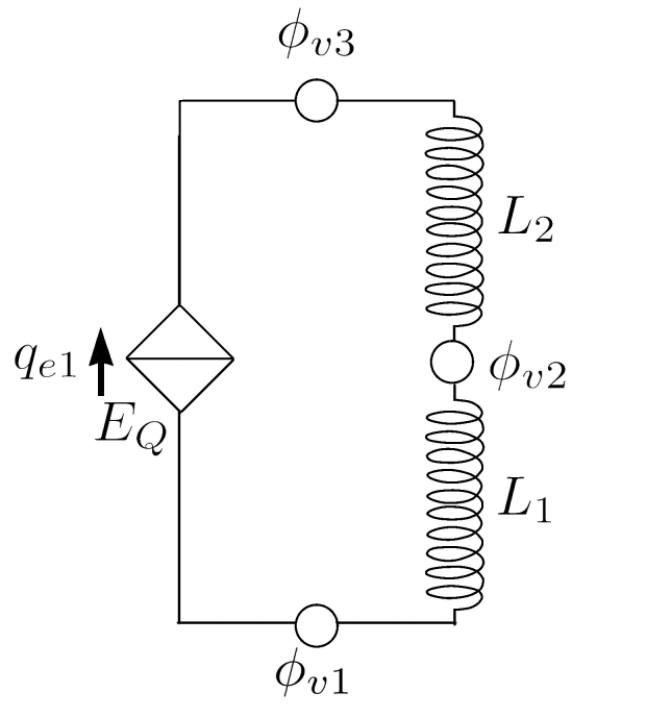}
         \caption{~}
         \label{fig1}
     \end{subfigure}
     \hfill
     \begin{subfigure}[b]{0.25\textwidth}
         \centering
         \includegraphics[width=\textwidth]{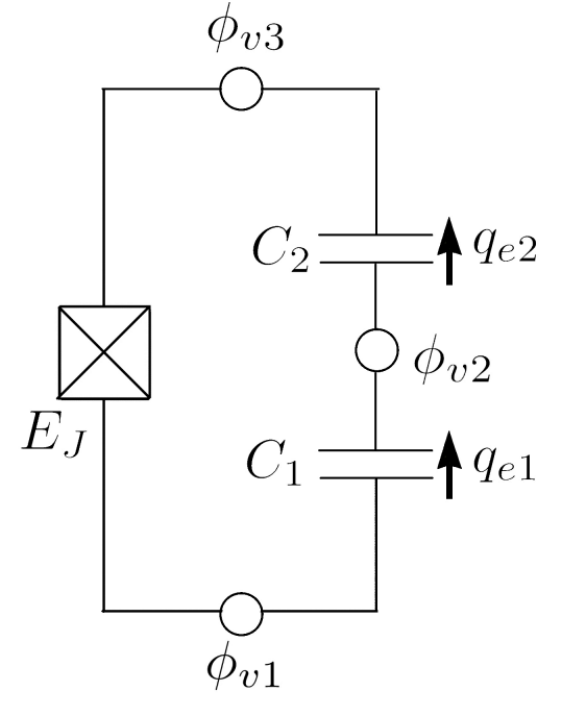}
         \caption{~}
         \label{fig3}
     \end{subfigure}
     \hfill
     \begin{subfigure}[b]{0.3\textwidth}
         \centering
         \includegraphics[width=\textwidth]{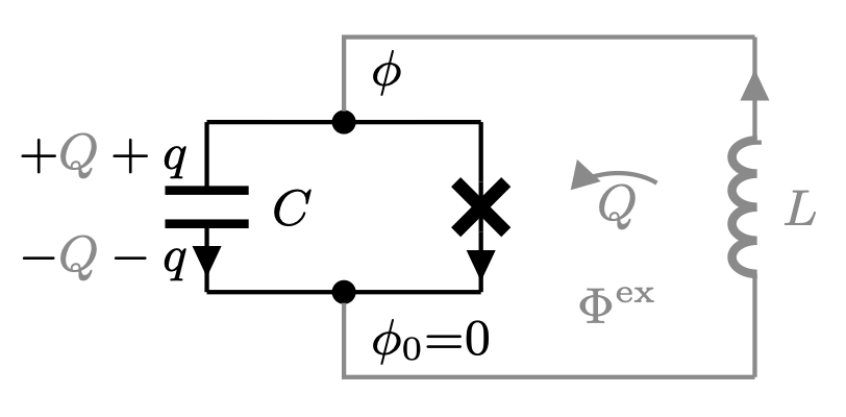}
         \caption{~}
         \label{fig4}
     \end{subfigure}
        \caption{(a) A circuit involving an inductively shunted island at $\phi_{2}$ \cite{sym}. (b) A circuit with a capacitive shunted island at $\phi_{2}$ \cite{sym}. (c) Idealized fluxonium circuit \cite{dual}. Here $\phi_{vi} \equiv \phi_i$ and $q_{ej}= q_j$, $v$ and $e$ correspond to the fact that the dofs belong to vertex and circuit element respectively.}
        \label{figs}
\end{figure}

{\bf{DCA of Inductively Shunted SQC:}} Construction of the Lagrangian corresponding to a specific SQC, following \cite{sym}, is briefly explained in Supplemental Material B.
We borrow a model from \cite{sym} with  corresponding  SQC in Figure (\ref{fig1}), 

\begin{equation}
L=q(\dot{\phi}_3-\dot{\phi}_1) +Ecos(\frac{2\pi q}{2e})-\frac{1}{2L_1}(\phi_1-\phi_2)^2 -\frac{1}{2L_2}(\phi_2-\phi_3)^2
    \label{1}
\end{equation}
In the terminology of \cite{sym} $\phi_i$ are three flux node variables and $q$ is the branch charge.
\begin{equation}
P_i=\frac{\partial L}{\partial \dot{\phi}_i}~,~i=1,2,3,~~ \pi =\frac{\partial L}{\partial \dot{q}}~;~\{\phi_i,P_{j}\}=\delta_{ij},~~\{q,\pi\}=1.
    \label{2}
\end{equation}
 The constraints are $\chi_1=P_1+q\approx 0,~\chi_2=P_2\approx 0,~\chi_3=P_3-q\approx 0,~\chi_4=\pi\approx 0$, but we use an equivalent linear combination 
\begin{equation}
\chi_1=P_1+P_3\approx 0,~\chi_2=P_2\approx 0,~\chi_3=P_3-q\approx 0,~\chi_4=\pi\approx 0 .
    \label{3a}
\end{equation}
The extended Hamiltonian is
\begin{equation}
H_E=-Ecos(\frac{2\pi P_3}{2e})+\frac{1}{2L_1}(\phi_1-\phi_2)^2 +\frac{1}{2L_2}(\phi_2-\phi_3)^2 +\sum_{i=1}^4 \alpha_i\chi_i .
    \label{5}
\end{equation}
 Time persistence $\dot{\chi_i}=\{\chi_i,H_{E}\}\approx 0$ reproduces  $\alpha_3,\alpha_4$ from $\dot{\chi_3}=0, \dot{\chi_4}=0 $ respectively. Since the SCC pair $\chi_3,\chi_4$ commutes with the other constraints,   DBs generated by them  can be calculated in the first stage - a useful property of DCA. In the present case, this does not induce any change in the brackets of the variables of interest.

Time persistence of the remaining constraints does not determine   $\alpha_1,\alpha_2$ but  generates a new constraint 
\begin{equation}
\dot {\chi}_1=\{\chi_1,H\}=((L_1+L_2)\phi_2-(L_1\phi_3+L_2\phi_1))/(L_1L_2)\equiv \chi_5
\approx 0, ~ \dot {\chi}_2=-\equiv \chi_5\approx 0 .
    \label{8}
\end{equation}
It can be shown  that $\Psi=\chi_1+\chi_2=P_1+P_2+P_3 \approx 0$ is a FCC since it commutes with rest of the constraints $\chi_1,\chi_2,\chi_5$. Since $\dot \Psi =\{\Psi,H\}\approx 0$ the chain of constraints stops and there are no further constraints and
\begin{equation}
H=-Ecos(\frac{2\pi P_3}{2e})+\frac{1}{2L_1}(\phi_1-\phi_2)^2 +\frac{1}{2L_2}(\phi_2-\phi_3)^2.
    \label{7}
\end{equation}
Since $\Psi$ is already considered as an FCC we can choose $\chi_5$ and either of $\chi_1,\chi_2$ as the SCC pair; we choose $\chi_2,\chi_5$ and with DBs in mind, use them strongly to obtain 
\begin{equation}
H=-Ecos(\frac{2\pi P_3}{2e})+\frac{1}{2(L_1+L_2)}(\phi_1-\phi_3)^2
    \label{10}
\end{equation}
The brackets between $P_3,\phi_1,\phi_3$ are not affected. Notice that $\{P_3,\phi_1-\phi_3\}=1$ is a canonical pair. This agrees with our DOF count since there are overall four SCC and one FCC $\rightarrow~ 8-(1\times 2+4\times 1)=2$ DOF in phase space or one DOF in configuration space. 

{\it{Additional freedom revealed in Dirac analysis:}} We still have the FCC $\psi\approx 0$ for which we can choose a gauge constraint $\Theta=\phi_1\approx 0$, (such that $(\psi,\Theta)$ are SCC), leading to
\begin{equation}
H=-Ecos(\frac{2\pi P_3}{2e})+\frac{1}{2(L_1+L_2)}(\phi_3)^2 .
    \label{11}
    \end{equation}
    Thus using DCA  we have recovered, much more economically, all the results of \cite{sym}, with an additional insight of a gauge invariance.

    As advertised earlier, we now show the flexibility induced by the FCC; a generalized  gauge condition $\Theta=a\phi_1+b\phi_3,~a,b \text{ arbitrary constants}$ will generate  different 
 DBs
\begin{equation}
\{\phi_1,P_1\}_{DB}=\frac{b}{a+b},~\{\phi_1,P_3\}_{DB}=-\frac{b}{a+b},
\{\phi_3,P_3\}_{DB}=\frac{a}{a+b},~\{\phi_3,P_1\}_{DB}=-\frac{a}{a+b} .
    \label{13}
    \end{equation}
    Using the gauge condition  $\phi_1=-\frac{b}{a}\phi_3$ we get $H$ and equations of motion
\begin{equation}
H=-Ecos(\frac{2\pi P_3}{2e})+\frac{1}{2(L_1+L_2)}(\frac{a+b}{a})^2(\phi_3)^2
    \label{14}
\end{equation}	
\begin{equation}
\dot{\phi}_3=\frac{a}{a+b}\frac{2\pi E}{2e}sin (\frac{2\pi P_3}{2e}),~\dot{P}_3=-\frac{a+b}{a}\frac{\phi_3}{L_1+L_2}
    \label{15}
\end{equation}	 
finally yielding
\begin{equation}
\ddot{\phi}_3=-(\frac{2\pi}{2e})^2 \frac{E}{L_1+L_2}cos (\frac{2\pi P_3}{2e})\phi_3.
    \label{16}
\end{equation}	
Note that (\ref{16}) it is independent of $a,b$ and so gauge invariant. Also notice that the bracket
$\{\phi_3,P_3\}_{DB}=\frac{a}{a+b}$ can be directly identified from the Lagrangian
\begin{equation}
L=\frac{a+b}{a}\dot{\phi}_3P_3 +Ecos(\frac{2\pi P_3}{2e})+\frac{1}{2(L_1+L_2)}(\frac{a+b}{a})^2(\phi_3)^2
    \label{ll}
\end{equation}	
and in fact with a redefinition $\frac{a+b}{a}\phi\rightarrow \phi$ the system becomes trivially canonical.\\
{\bf{DCA of  Capacitively Shunted SQC:}} Contrary to  \cite{sym}, that invoked different procedures to study (right null vectors of the symplectic form for) inductively shunted  and (Noether charge for) capacitatively shunted   models, {\it{we use the same DCA to solve the latter}}, in Figure (\ref{fig3}).  The corresponding Lagrangian is \cite{sym}
\begin{equation}
    L = q_1(\Dot{\phi}_{2}-\Dot{\phi}_{1}) + q_{2}(\Dot{\phi}_{3}-\Dot{\phi}_{2}) + E\cos{\left( 2 \pi \frac{\phi_{1}- \phi_{3}}{\phi_0} \right)}- \frac{1}{2 C_1} q_{1}^2 - \frac{1}{2 C_2} q_2^2 .
\end{equation}
With $
    p_i = \frac{\partial L}{\partial \Dot{\phi}_{i}}, P_i = \frac{\partial L}{\partial \Dot{q}_{i}}$ With $\{\phi_{i},p_{j}\}=\delta_{ij},~\{q_i,P_j\}=\delta_{ij}$. 
The constraints $\chi_k\approx 0$ are 
\begin{equation}
    \chi_1 = p_1 + q_1,~
    \chi_2 = p_3 - q_2,~
    \chi_3 = p_2 - q_1 + q_2,~
    \chi_4 = P_1,~ 
    \chi_5 = P_2 .
    \label{p1}
\end{equation}
 Now $\Psi=\chi_1+\chi_2+\chi_3=p_1+p_2+p_3\approx 0$  is an FCC. Of the remaining four independent SCC,  $(\chi_1, \chi_4)$ and $(\chi_2, \chi_5)$ are mutually commuting SCC pairs; DBs can be constructed for the individual pairs separately. There are four SCC and one FCC with $\phi_1,\phi_2,\phi_3,q_1,q_2$ "coordinates" and their momenta. Number of independent DOF is $5\times 2 -(4\times 1 +1\times 2)=4$; there are two full DOF in configuration space.

Hamiltonian on constraint surface and DBs are
\begin{equation}
    H = - Ecos\left(\frac{2 \pi}{\phi_0}(\phi_{1} - \phi_{3})\right) + \frac{1}{2C_1}q_1 ^2 + \frac{1}{2C_2}q_2 ^2,~~
    \{\phi_1,q_1\}=1,~~\{\phi_3,q_2\}=1.
   \label{p3} 
\end{equation}
 Dynamics is given by
\begin{equation}
 \dot{\phi}_1=\frac{q_1}{C_1} ,~ \dot{q}_1=-\frac{2\pi E}{\phi_0}  sin \left(\frac{2 \pi}{\phi_0}(\phi_{1} - \phi_{3})\right),~~ 
  \dot{\phi}_3=\frac{q_2}{C_2} ,~ \dot{q}_2=+\frac{2\pi E}{\phi_0}  sin \left(\frac{2 \pi}{\phi_0}(\phi_{1} - \phi_{3})\right) .
  \label{p4}
\end{equation}
Defining $\phi_1-\phi_3=Q$ we find $
    \ddot {Q}=-(\frac{1}{C_1}+\frac{1}{C_2})\frac{2\pi E}{\phi_0}  sin \left(\frac{2 \pi}{\phi_0}Q\right)$.
Note that $d(q_1+q_2)/dt=0 \rightarrow ~q_1+q_2$ is a conserved quantity and so the corresponding cyclic "coordinate" proportional to $\phi_1+\phi_3$ is absent. The equivalent circuit is a SQC with $C_1$ and $C_2$ in series combination. Once again the DCA manages to recover correct results as  \cite{sym}.\\
{\bf{DCA of a "Noncommutative" SQC :}}
Consider the circuit in Figure (\ref{fig4}) and the corresponding Lagrangian, discussed in \cite{dual}
\begin{equation}
L=\frac{C}{2}\dot{\phi}^2 +\frac{L}{2}\dot{q}^2 +Ecos(\frac{2\pi \phi}{A})-q\dot{\phi}+\dot{q}\Phi^{ext} 
\label{n1}
\end{equation}
with $\Psi$ an external source, $\phi_1$  is the flux associated with Cooper pairs and $\phi_2$ is the loop charge \cite{dual}.
In \cite{dual}  a dual approach using loop charge representation to quantize the SQC is applied but in DCA it is an easily solvable simple model without constraints. 

Instead we  solve  a slightly different model by assuming small $\dot{\phi}_2 \rightarrow\dot{\phi}^2_2\approx 0$ leading to
\begin{equation}
L=\frac{C}{2}\dot{\phi}^2 + Ecos(\frac{2\pi \phi}{A})-q\dot{\phi}+\dot{q}\Phi^{ext}
\label{n2}
\end{equation}
Our aim is to show  that the resulting constraint system has a non-zero DB between $\phi$ and $q$, which in this case are not a conjugate pair, but rather like "coordinates" thus inducing a {\it{"spatial" non-commutativity}} in  the system \textbf{---} an analogue of spatial non-commutative quantum mechanics \cite{noncom}. To motivate this modification we point out that  the issue of noncommutativity is important in High energy Physics and this specifically designed SQC can act as a novel {\it{analogue model}.}

The Hamiltonian 
\begin{equation}
H=\frac{1}{2C}(P_1+q)^2-Ecos(\frac{2\pi \phi}{A})
\label{n3}
\end{equation}
 with  momenta $P_1=C\dot{\phi}-q,~P_2=\Psi$, has two SCCs  $\chi_1=P_2-\Psi\approx 0, ~~ {\chi}_2=\dot{\chi}_1=P_1+q\approx 0.$
With $\{\phi,q\}_{DB}=-1,~\{\phi,P_1\}_{DB}=1,~H=-Ecos(\frac{2\pi \phi}{A})$ on constraint surface
 equations of motion are
\begin{equation}
\dot{\phi}=\{\phi,H\}=0,~\dot{q=\{q,H\}=-\frac{2\pi E}{A}sin(\frac{2\pi \phi}{A})}
\label{nn8}
\end{equation}
so that the first relation in (\ref{nn8}) gives $\phi=C$, with $C$ a constant, and $\dot{q}=-\frac{2\pi E}{A}sin(\frac{2\pi \phi}{A})|_{\phi=C}=-\frac{2\pi E}{A}sin(\frac{2\pi C}{A})$, another constant.

Interestingly the symplectic structure (or kinetic term) in the action $q \dot{\phi}$ appears in condensed matter physics in well known Landau problem, i.e. a charged particle moving in $\phi_1-\phi_2$-plane, in a constant magnetic field perpendicular to the plane, with the condition that the magnetic field is large so that  kinetic energy of the particle is negligible. We have reproduced a circuit analogue of  a well-known  non-commutative quantum mechanical problem \cite{noncom}.\\
{\bf{DCA of a generic SQC:}} Our aim is to show that the DCA approach is sufficient to indicate that "noncommutativity in SQC, presented earlier, is quite common and can appear in a wide class of SQCs governed by a generic Lagrangian (here we do not consider is circuit implementation),
\begin{equation}
L=q(\dot{\phi}_3-\dot{\phi}_1) +E  cos(\frac{2\pi q}{2e})-\frac{1}{2L_1}(\phi_1-\phi_2)^2 -\frac{1}{2L_2}(\phi_2-\phi_3)^2 +\lambda_1\dot{\phi}_1\phi_3 +\lambda_2\dot{\phi}_1\phi_2 +\lambda_3\dot{\phi}_2\phi_3 .
    \label{nn1}
\end{equation}
The constraints $
\chi_1=P_1+P_3-\lambda_1\phi_3-\lambda_2\phi_2\approx 0,~\chi_2=P_2-\lambda_3\phi_3\approx 0,~\chi_3=P_3-q\approx 0,~\chi_4=\pi\approx 0 $ and 
 extended Hamiltonian
\begin{equation}
H=-Ecos(\frac{2\pi P_3}{2e})+\frac{1}{2L_1}(\phi_1-\phi_2)^2 +\frac{1}{2L_2}(\phi_2-\phi_3)^2 +\sum_{i=1}^4 \alpha_i\chi_i
    \label{nn5}
\end{equation}
reveal  that  $\dot{\chi}_i\approx 0$ fixes all the $\alpha_i$, so there are no further constraints. The constraint matrix $\{\chi_i,\chi_j\}$ is non-singular; $\chi_i$s comprise a set of four SCC and  no FCC as in  previous cases. We will have $8-(4\times 1)=4$ DOF in phase space or two independent DOF in configuration space.

The full set of DBs among "coordinate" variables are given by 

\begin{equation}
\begin{split}
\{\phi_1,\phi_2\}_{DB}=\frac{1}{\lambda_2-\lambda_3},~\{\phi_1,\phi_3\}_{DB}=0,~\{\phi_1,q\}_{DB}=\frac{\lambda_3}{\lambda_2-\lambda_3}, \\
\{\phi_2,\phi_3\}_{DB}=-\frac{1}{\lambda_2-\lambda_3},~\{\phi_2,q\}_{DB}=-\frac{\lambda_1}{\lambda_2-\lambda_3},~\{\phi_3,q\}_{DB}=\frac{\lambda_2}{\lambda_2-\lambda_3}
    \label{n6}
\end{split}
\end{equation}	

 The significance of noncommutative quantum circuits is now clear since conventionally noncommutative quantum mechanics refers to models where spatial coordinates do not commute.  Rest of the DBs are can also be computed in a straightforward way but for the present model these are not required. We can exploit the SCCs $\chi_i=0$ as strong relations to express the Hamiltonian as
\begin{equation}
H=-Ecos(\frac{2\pi q}{2e})+\frac{1}{2L_1}(\phi_1-\phi_2)^2 +\frac{1}{2L_2}(\phi_2-\phi_3)^2 +\sum_{i=1}^4 \alpha_i\chi_i .
    \label{nn7}
\end{equation}
It is interesting to note that the under DB the DOFs $(Q_1,P_1)$ and $(Q_2,P_2)$  constitute {\it{two independent canonical degrees of freedom}} where 
\begin{equation}
 Q_1=(\phi_3-\phi_1),~P_1=q, ~Q_2=\phi_2+\frac{\lambda_1}{\lambda_3}\phi_1,~P_2=-\lambda_2\phi_1+\lambda_3\phi_3 ,
     \label{n8}
\end{equation}	
with the only non-vanishing DBs being $\{Q_1,P_1\}_{DB}=1,~\{Q_2,P_2\}_{DB}=1$. This result agrees with our advertised DOF count. 

Let us invert the relations (\ref{n8})
\begin{equation}
\phi_1=\frac{\lambda_3-P_2}{\lambda_2-\lambda_3},~\phi_2=\frac{Q_2\lambda_3(\lambda_2-\lambda_3)-\lambda_1(\lambda_3Q_1-P_2)}{\lambda_3(\lambda_2-\lambda_3)},~\phi_3=\frac{Q_1\lambda_2-P_2}{\lambda_2-\lambda_3},~q=P_1 .
    \label{n9}
\end{equation}	
Keeping the unique noncommutative feature of the circuit intact let us consider a simplified model with $\lambda_1=\lambda_2=0,~\lambda_3=\lambda$ so that the Dirac algebra simplifies to
\begin{equation}
\{\phi_1,\phi_2\}_{DB}=-\frac{1}{\lambda},~\{\phi_1,\phi_3\}_{DB}=0,~\{\phi_1,q\}_{DB}=-1,~
\{\phi_2,\phi_3\}_{DB}=\frac{1}{\lambda}, $$$$
\{\phi_2,q\}_{DB}=0,~\{\phi_3,q\}_{DB}=0 .
    \label{n6a}
\end{equation}	
The Hamiltonian  is expressed in terms of canonical variables as 
\begin{equation}
H=-Ecos(\frac{2\pi P_1}{2e})+\frac{1}{2L_1 \lambda^2}(\lambda (Q_1+Q_2)-P_2)^2 +\frac{1}{2L_2\lambda^2}(\lambda Q_2-P_2)^2  .
    \label{n10}
\end{equation}	 
This will yield the equations of motion
\begin{equation}
\ddot{Q}_1=(\frac{2\pi}{2e})^2Ecos(\frac{2\pi P_1}{2e})(-\frac{Q_1+Q_2}{L_1}+\frac{P_2}{\lambda L_1}), ~
\ddot{Q}_2=-\frac{1}{L_2\lambda^3}(\frac{\lambda}{L_1}((Q_1+Q_2)-P_2).
    \label{n11}
\end{equation}	
Notice that for a low momentum approximation $P_1\approx 0,~P_2\approx 0$, the system reduces to a single simple harmonic oscillator, (for $Q_1+Q_2=Q$),
\begin{equation}
\ddot Q=-\left  ((\frac{2\pi}{2e})^2\frac{E}{L_1}+\frac{1}{L_1L_2\lambda^3}\right )Q .
    \label{n12}
\end{equation}	
{\bf{Circuit Quantisation:}}
So far the examples studied in the present work and the brief discussion of Dirac's Hamiltonian analysis of constraint systems pertains to classical systems. But the true power and universality of Dirac's formalism manifests itself in {\it{quantization of constraint systems}}. An essential reason for this is that Dirac's scheme is based on Hamiltonian formalism; a generalized (Poisson-like or Dirac) bracket structure that can be directly exploited in transition to quantum regime. In this scheme a generalization of the Correspondence Principle is proposed \cite{dir}; in case of constraint systems the Dirac Brackets have to be elevated to the status of quantum commutators,
\begin{equation}
    \{A, B\}_{DB}=C ~~\rightarrow ~~[\hat A,\hat B]=i\hbar \hat C
    \label{d1}
\end{equation}
 In some models (in particular where the constraints are non-linear functions of the dynamical variables), $C$ can depend on  the dynamical variables as well leading to operator valued $\hat C$ upon quantization. This can give rise to  complications but there are systematic ways of treating such problems. But in the SQCs considered in this work, that happens to be the generic cases in this context, $C$ in (\ref{d1}) is a constant (matrix) that can easily be scaled away to yield a canonical form.

In majority of the examples discussed above, one ends up with a harmonic oscillator-like system that is elementary to solve as a quantum problem. As discussed in \cite{sqed}, it is quite feasible to have quantum effects manifested in this type of macroscopic systems under appropriate conditions.

{\bf{Conclusion and future outlook:}} 
From surveying recent literature it appears that a unified framework is required in the  theoretical analysis  of Superconducting Quantum Circuits, a key element in quantum information and quantum computing technology. Qualitatively distinct schemes are exploited for different forms of SQCs (involving sophisticated mathematical tools) that generate correct results. Indeed it would be desirable if a systematic and conceptually clear formalism is found that can address all types of SQCs in equal footing and yield    the effective canonical degrees of freedom, that have to be quantized eventually. This issue is non-trivial because in the Lagrangian approach, complications can arise due to the presence of constraints. \\  
In the present work we have conclusively established that the Hamiltonian framework of constraint dynamics, as formulated by Dirac, is highly appropriate to analyse Superconducting Quantum Circuits. Under this scheme, all types of such SQCs discussed in literature, can be addressed in a unified and unambiguous way. We have considered physically realised SQCs and also SQC in a {\it{gedanken}} experiment scenario. Apart from reproducing the existing results (such as canonical variables and the correct Hamiltonian)  with minimal computation, our results go further in revealing the symmetries of the SQCs thereby introducing more flexibility in choosing equivalent canonical degrees of freedom. Specifically we have discussed in detail the (conventional) inductively and capacitatively shunted SQCs  and more complex arbitrary form of SQC, whose equivalent circuit needs to be achieved. Furthermore, from a quantum mechanics perspective, we have discussed an SQC, existing in literature, that in certain limit can represent noncommutative quantum mechanics where the coordinates do not commute.  

Finally we would like to comment on a recent work \cite{mart} that discusses "failure" of DCA in case of  nearly singular SQC; in particular examples of SQC with very low capacitance (effectively playing the role of a small pass particle in a mechanical circuit) are suggested. It needs to be appreciated that the difference between very small "mass" and exactly zero "mass" can be critical as it can very well introduce (or remove as the case may be) symmetries of the system, that will be reflected in the nature of constraint classification, changing Second Class to First Class Constraint (or {\it{vice versa}}). For example in (\ref{d5}), if the constraint matrix $\chi_{kl}=function ~of ~"mass"$, it might so happen that the RHS vanishes for $"mass"=0$ so that $\chi_{kl}$ becomes singular, indicating that some of the Second Class Constraints are now First Class, leading to additional symmetries (and maybe new degrees of freedom). Obviously one will get nonsensical results if one uses DCA and tries to  obtain $"mass"=0$ results by considering the zero "mass" limit of $\chi_{kl}=function ~of ~"mass"$. DCA demands, in this case, that  non-zero "mass" and zero "mass" models are to be treated individually, and not as a zero "mass" limit from former to latter model. Furthermore, it needs to be stressed that in \cite{mart} problems appear when the system Hamiltonian becomes multiple valued such that the model itself is pathological; the failure is not a weakness of DCA.   Hence  the claim is that given the Lagrangian for any SQC, however complicated, the method presented here, can be exploited in solving the circuit dynamics. On the other hand, going in the reverse direction, one can posit a Lagrangian with certain preconceived featured, solve the model in Dirac framework and subsequently visualize the corresponding SQC.

To conclude, in the present work we have conclusively established that Dirac's Constraint Analysis (in Hamiltonian framework) is a universal platform to study different types of Superconducting Quantum Circuits in a unified way. This is accomplished by correctly solving well known SQCs in the Dirac approach. It should be stressed that in previous works, each of these SQCs were treated in distinct schemes, involving quite sophisticated mathematical concepts, which however are redundant as shown here. Lastly we have also suggested a new model, a generic form of SQC, that can also be addressed properly in the Dirac formalism. It would be interesting to compare these new results with a circuit implementation of the same. 
\\ 

\large{\bf{Supplemental Material}}

{\bf{Supplemental Material A:}} We briefly explain the study of constraint systems in a Hamiltonian formalism, as pioneered by Dirac \cite{dir}.  A   singular Lagrangian $L(q_i,\dot{q}_i)$ is defined as  one for velocities $\dot {q}_i$ can not be expressed uniquely in terms of canonical momenta $p^i=\frac{\partial L}{\partial \dot{q}_i}$, due to the presence of constraints, relations between coordinates and momenta only.

For non-singular $L$, the Poisson Bracket (PB) $\{q_i,p^j\}=\delta^i_j$ and canonical Hamiltonian 
\begin{equation}
H_c=p^i\dot q_i-L
    \label{a1}
\end{equation}
generates the equations of motion,
\begin{equation}
\dot q_i=\{q_i,H_c\}=\frac{\partial H_c}{\partial {p}_i},~\dot p^i=\{p^i,H_c\}=-\frac{\partial H_c}{\partial {q}_i} .
    \label{a2}
    \end{equation}
A necessary and sufficient condition for $L$ to be   singular is 
\begin{equation}
det|\frac{\partial^2 L}{\partial \dot{q}_i \partial \dot{q}_i} =0.
    \label{a3}
\end{equation}
This indicates that there are {\it{primary constraints}}
\begin{equation}
\phi_m(q_i,p^i)\approx 0
    \label{a4}
\end{equation}
where $\approx$ indicates a {\it{weak equality}} meaning that some dynamical variables can have non-zero PB with $\phi_m(q_i,p^i)$. Another way of saying is that {\it{weak}} equations are satisfied only on the constraint surface (when the constraints are applied) and can be invoked after the requisite PBs are computed. On the other hand {\it{strong}} equations, involving {\it{strong equality}} are always true, independent of constraints.  Thus the Hamiltonian is not unique and one introduces an extended Hamiltonian $\tilde H$
\begin{equation}
\tilde H_e=H_c+u_m\phi_m \approx H_c.
    \label{a5}
\end{equation}
$H_e$ is weakly equal to $H_c$ and $u_m$ are so far arbitrary multiplier fields.  Thus the general form of equations of motion are
\begin{equation}
\dot q_i\approx\{q_i,H_e\}=\frac{\partial H_c}{\partial {p}_i} +\frac{\partial \phi_m}{\partial {p}_i}u_m , ~ 
\dot p_i\approx\{q_i,H_e\}=-\frac{\partial H_c}{\partial {q}_i} -\frac{\partial \phi_m}{\partial {p}_i}u_m .
    \label{a2}
    \end{equation}
 Preserving the constraints $\phi_m$ for all time requires
\begin{equation}
\dot{\phi}_m = \{\phi_m,H_e\}=\{\phi_m,H_c\}+\{\phi_m,\phi_n\}u_n\approx 0 .
    \label{a3}
\end{equation}
From (\ref{a3}), either some of the $u_n$s can get fixed or new relations, independent of $u_n$s and linearly independent from previous {\it{primary constraints}} $\phi_m$ can arise. These are termed as {\it{secondary constraints}} and they are to be appended to the set of {\it{primary constraints}}. These process has to be continued to generate (if present) {\it{tertiary constraints}} and so on until all the $u_m$s are determined and the full set of constraints $\phi_\alpha$ (including {\it{primary, secondary, .... constraints}}) are obtained satisfying
\begin{equation}
\tilde H_e=H_c+u_m(q,p)\phi_m(q,p), ~~\{\phi_\alpha,H_e\}\approx 0.
    \label{a4}
\end{equation}
Now comes the important classification of constraints. In general a {\it{first class}} quantity $R(q,p)$ is defined to satisfy
\begin{equation}
\{R,\phi_\alpha\}\approx 0
 \label{a5}   
\end{equation}
and quantities $Z(q,p)$ not satisfying the above are termed as {\it{second class}} quantity
\begin{equation}
\{R,\phi_\alpha\}\neq 0 .
 \label{a6}   
\end{equation}
The complete set of constraints $\phi_\alpha$ can be divided in to sets of {\it{first class}} constraints $\psi_a$ satisfying 
\begin{equation}
\{\psi_a,\phi_\alpha\}\approx 0
 \label{a7}   
\end{equation}
and rest of the constraints $\phi_k$ not satisfying the above being {\it{second class}} constraints
\begin{equation}
\{\phi_k,\phi_l\}\neq 0 .
 \label{a8}   
\end{equation}
As Dirac has shown the PBs between {\it{second class}} constraints $\phi_k$ generate a non-singular (or invertible) matrix $C_{kl}$
\begin{equation}
\{\phi_k,\phi_l\}=C_{kl},~~C_{kl}{C}^{-1}lm=\delta_{km} .
 \label{a9}   
\end{equation}
Since the constraint matrix $C_{kl}$ is antisymmetric by construction and non-singular it has to be of even dimensions indicating that there can be only even number of {\it{second class}} constraints in s system.

This leads to the definition of Dirac Bracket (DB) between two arbitrary quantities $A,B$ as 
\begin{equation}
\{A,B\}_{DB}=\{A,B\}-\{A,\phi_k\}{C}^{-1}km\{\phi_m,B\}
 \label{a10}   
\end{equation}
with the essential property that
\begin{equation}
\{A,\phi_k\}_{DB}=\{\phi_k,A\}_{DB}=0
 \label{a11}   
\end{equation}
that is {\it{Dirac brackets between any quantity and the {\it{second class}} constraints strongly vanish}}. This allows us to invoke the {\it{second class}} constraints as strong relations $\phi_k(q,p)=0$ and use each one to eliminate one degree of freedom (or dynamical variable) in phase space, keeping in mind that in all subsequent operations PBs have to be replaced by Dirac Brackets. Dirac Brackets satisfy the properties obeyed by the PBs, they are antisymmetric, obey  chain rule and satisfy Jacobi identity.

Thus we are left with the set of {\it{first class}} constraints $\psi_a$, Dirac Brackets between the remaining variables (after eliminating some using the {\it{second class}} constraints $\psi_k$ and the total Hamiltonian $H_t$
\begin{equation}
H_t=H' + v_a\psi_a 
 \label{12}   
\end{equation}
where $v_a$ are arbitrary and $H'=H_c-\{H_c,\phi_k\}C^{-1}_{kl}\phi_l$. Time evolution of a quantity $A$ will be given by 
\begin{equation}
\dot A=\{A,H_t\}_{DB}
 \label{13}   
\end{equation}
which will contain some amount of arbitrariness due to the explicit presence of $v_i$. 

Presence of a {\it{first class}}  constraint $\psi_a$ indicates gauge invariance of the system meaning that the gauge transformation generated by $\psi_a$, $\delta q=\{q,\psi_a\}~_{DB},\delta p=\{p,\psi_a\}_{DB} $ do not affect the physical state (and hence physics) of the system. This is reflected in the presence of arbitrariness through $v_a$ in the time evolution since the original Lagrangian contained gauge degrees of freedom. It is possible to remove all arbitrariness by introducing  {\it{gauge fixing condition}} $\chi_a$ for each {\it{first class}} constraint $\psi_a$ such that $\psi_a,\chi_a$ form a pair of {\it{second class}} constraints, $\{\psi_a,\chi_a\}_{DB}\neq 0$ which will generate a new set of Dirac Brackets following the same procedure. The {\it{gauge fixing condition}} can be chosen as per convenience and it is not dictated by the Lagrangian. In this way if {\it{gauge fixing conditions}} are introduced to fix all the gauge invariances (i.e.  number of independent $\chi_a$ equal to number of $\phi_a$, we can use $\chi_a=0,~\phi_a=0$ as strong relations, remove once again a degree of freedom for each  {\it{second class}} constraint of this new set, we will finally end up with a Hamiltonian comprising of a reduced number of degrees of freedom, (that are physical in nature), and their time evolution will be governed by the final set of Dirac Brackets, without any arbitrariness in the system.

From the above discussion we find that each Second Class Constraint can be strongly used to remove one degree of freedom in phase space and each First Class Constraint together with its gauge fixing constraint can remove two degrees of freedom in phase space. The dimension of the reduced phase space (number of independent and physical degrees of freedom in phase space) after considering Dirac Brackets for all Second Class Constraints and fixing gauge condition for all First Class Constraints  will be given by \\
{\it{Number of physical degrees of freedom = (total number of degrees of freedom) - (total number of Second Class Constraints) - 2$\times $ (total number of First Class Constraints)}}.

Dirac's proposal was to quantise this final Hamiltonian system by elevating the Dirac Brackets to quantum commutators.

In our paper in the main text we have precisely exploited this procedure.

For the sake of completeness, (although not relevant for our work), we mention that in  a specific problem, implementing the quantisation programme by converting Dirac Brackets to quantum commutators might give rise to computational problems (such as operator ordering, ..) since the right hand side of Dirac Brackets can be non-constant and depend on the degrees of freedom involved. Sophisticated techniques have been developed to overcome such contingencies which , however are beyond the scope of the present work.\\
\vskip .5cm
{\bf{Supplemental Material B:}} As a simple example, consider an inductor $L$ and a capacitors $C$ there
is a single degree of freedom  flux $x$  across the inductor. The corresponding  Lagrangian, interpreted as an effective kinetic energy minus potential energy form is given by  
\begin{equation}
L=\frac{1}{2}C\dot {x}^2 - \frac{1}{2L} {x}^2 .
 \label{14}   
\end{equation}
Defining the canonical momentum $X=\frac{\partial L}{\partial \dot x}$ the Hamiltonian is given by
\begin{equation}
H=\frac{1}{2C}X^2 + \frac{1}{2L}x^2.
 \label{15}   
\end{equation}
As explained in detail in \cite{sqc}, the construction of the action  involves introducing   charge variables $X_i$ passing through capacitative branches  and flux variables $x_i$ on nodes the capacitive subgraph of the
circuit as conjugate degrees of freedom (see figure).  Time integrals of  voltages and currents in a circuit yield  flux $x_i$  and electric charge $X_i$ respectively.  Kirchhoff’s voltage
law states  that the sum of the voltage drops around a loop vanishes in the absence of external magnetic fields. This suggests the assignment of  voltages on nodes (or vertices)  of the circuit. On the other hand, it is natural to  define currents on branches (or edges)  in the circuit. Similar to position and momentum in mechanical circuits, $x_i,X_i$ act as canonically conjugate variables.  Indeed, a generic 
 circuit can have  multiple elements and therefore multiple flux and charge variables. {\it{ In general, there is not a one-to-one mapping between the branches  and nodes  of a
circuit}}. Clearly this feature translates into a mismatch between the number of coordinates and momenta in a mechanical circuit, which is nothing but an example of a constraint system. In the paper this type of constraint circuits are analysed in the framework of Dirac's formalism \cite{dir}.

{\bf{Data Availability Statement: No Data associated in the manuscript}}

\end{document}